\documentclass[12pt]{article}
\usepackage{graphicx}
\usepackage{psfig,picinpar,floatflt,amssymb,epsf}
\csname @addtoreset\endcsname{equation}{section}

\def\bseq{\begin{subequation}}  
\def\eseq{\end{subequation}}
\def\bsea{\begin{subeqnarray}}  
\def\esea{\end{subeqnarray}}

\newcommand{\bbox}{\lower.2ex\hbox{$\Box$}}

\newcommand{\beq}{\begin{equation}}
\newcommand{\eeq}{\end{equation}}
\newcommand{\bea}{\begin{eqnarray}}
\newcommand{\eea}{\end{eqnarray}}
\newcommand{\ena}{\end{eqnarray}}

\renewcommand{\a}{\alpha}
\renewcommand{\b}{\beta}

\renewcommand{\d}{\delta}
\renewcommand{\th}{\theta}
\newcommand{\thb}{{\bar{\theta}}}

\newcommand{\pa}{\partial}
\newcommand{\g}{\gamma}

\newcommand{\D}{\Delta}
\newcommand{\e}{\epsilon}

\renewcommand{\l}{\lambda}

\newcommand{\p}{\pi}

\renewcommand{\o}{\omega}

\newcommand{\Db}{\bar{D}}

\newcommand{\Phib}{\bar{\Phi}}

\newcommand{\qb}{\bar{q}}
\newcommand{\hb}{\bar{h}}

\newcommand{\ad}{{\dot{\alpha}}}

\begin{document}
\begin{titlepage}
\begin{flushright}
IFUM--842--FT \\
Bicocca--FT--05--18 \\
\end{flushright}
\vspace{.3cm}
\begin{center}
{\Large \bf Exact results in planar ${\cal{N}}=1$ superconformal \\
Yang--Mills theory }
\vfill
{\large \bf Andrea Mauri$^1$, Silvia Penati$^2$, Alberto Santambrogio$^2$ and
Daniela Zanon$^1$}\\
\vspace{0.6cm} 
{\small $^1$ Dipartimento di Fisica, Universit\`a di Milano and\\ INFN, Sezione di Milano, 
Via Celoria 16, I--20133 Milano, Italy\\
  $^2$ Dipartimento di Fisica, Universit\`a di Milano--Bicocca and\\ INFN, Sezione di 
Milano, Piazza della Scienza 3,
I--20126 Milano, Italy\\}
\end{center}
\vfill
\begin{center}
{\bf Abstract}
\end{center}
{\small  In the $\b$-deformed ${\cal N}=4$ supersymmetric $SU(N)$ Yang-Mills theory we study the 
class of operators ${\cal O}_{J}= {\rm Tr}(\Phi_i^J \Phi_k)$, $i\neq k$ and compute their exact anomalous 
dimensions for $N,J\rightarrow\infty$. This  leads to a prediction for the masses of the 
corresponding states in the dual string theory sector. We test the exact formula perturbatively up 
to two loops. The consistency of the perturbative calculation with the exact result indicates that 
in the planar limit the one--loop condition $g^2=h \overline{h}$ for superconformal invariance is 
indeed sufficient to insure the {\em exact} superconformal invariance of the theory. We present a 
direct proof of this point in perturbation theory. The ${\cal O}_{J}$ sector of this theory shares 
many similarities with the BMN sector of the ${\cal N}=4$ theory in the large R--charge limit.} 
\vspace{2mm} \vfill \hrule width 3.cm
\begin{flushleft}
e-mail: andrea.mauri@mi.infn.it\\
e-mail: silvia.penati@mib.infn.it\\
e-mail: alberto.santambrogio@mi.infn.it\\
e-mail: daniela.zanon@mi.infn.it
\end{flushleft}
\end{titlepage}

\section{Introduction}

The recent past has seen a new interest on exactly marginal deformations of 
${\cal N}=4$ SYM theory preserving ${\cal N}=1$ supersymmetry \cite{LS}, 
in particular after the supergravity duals of the so--called 
$\beta$-deformations of ${\cal N}=4$ $SU(N)$ SYM theory have been found by 
Lunin and Maldacena in \cite{LM}. 

Now new results start to emerge also from the field theory side: in \cite{FG,PSZ4,RSS} various 
properties of composite operators of the deformed theory have been investigated at the perturbative 
level (see also \cite{NP,MNSS,FRT1,FRT2}). The outcome is that the deformed theory shares some properties with 
the undeformed ${\cal N}=4$ theory, but new features emerge, as for example the finite corrections 
of the two- and three-point functions of protected operators \cite{PSZ4,RSS}.\\ The chiral ring of 
the theory was identified in \cite{BL,BJL} for generic values of the deformation parameter. It is 
given by the operators ${\rm Tr}(\Phi_i^J)$, $i=1,2,3$, and ${\rm Tr}(\Phi_1^J \Phi_2^J \Phi_3^J)$. 
In \cite{FG,PSZ4} it was shown that also the operator ${\rm Tr}(\Phi_1\Phi_2)$ does not acquire 
anomalous dimension.

In this paper we will focus on the operators ${\rm Tr}(\Phi_i^{J}\Phi_j )$, $i \neq j$. As opposed 
to what happens in the undeformed ${\cal N}=4$ case, these operators are not protected and their 
anomalous dimension was computed at one-loop in \cite{FG}. Our interest in these operators is 
motivated by the fact that in the large $J$ limit they resemble the BMN operators of the ${\cal 
N}=4$ theory \cite{BMN}. Indeed a perturbative superfield analysis performed at low orders shows 
that the supergraphs contributing to their anomalous dimension are exactly the same as the ones of 
the BMN case. Then we apply the derivation of \cite{SZ} to this class of operators and compute 
their exact anomalous dimension in the planar limit.

The consistency between the perturbative supergraph approach and the result obtained by using the 
method of \cite{SZ} suggests that the one-loop superconformal invariance condition remains valid in 
the planar limit to all orders of perturbation theory (at least for real values of the $\beta$ 
parameter of the $\beta$-deformation which is the case we consider here). We confirm this result by 
exploiting the formal analogy between $\beta$-deformed and non-commutative field theories.

The paper is organized as follows. In Section 2 we present the setup for the perturbative 
calculation that we perform in Section 3, where we compute the anomalous dimensions of the 
operators ${\rm Tr}(\Phi_i^{J}\Phi_j )$, $i \neq j$ in the large $N$ limit up to order 
$g^4$. The contributing graphs are the same that one would encounter in the calculation of the 
anomalous dimension of BMN operators. In Section 4 we apply to our operators the procedure 
introduced in \cite{SZ} and give an all-order result for their anomalous dimensions in the 
large $J$ and large $N$ limit. Then in 
Section 5 we prove that the one-loop condition of superconformal invariance remains valid to all 
orders in the planar limit.

\section{Generalities}

We consider the following deformation of the ${\cal N}=4$ SYM theory
\bea
S[j,\bar{j}]
&=&\int d^8z~ {\rm Tr}\left(e^{-gV} \Phib_i e^{gV} \Phi^i\right)+
\frac{1}{2g^2}
\int d^6z~ {\rm Tr} W^\a W_\a\nonumber\\
&&+ih  \int d^6z~ {\rm Tr}( q ~\Phi_1 \Phi_2 \Phi_3 - \frac{1}{q} \Phi_1 \Phi_3 \Phi_2 )
 - i \overline{h} \int d^6\bar{z}~ {\rm Tr} ( \qb ~\Phib_1 \Phib_3 \Phib_2 - \frac{1}{\qb}
\Phib_1 \Phib_2 \Phib_3 )
\nonumber\\
&&+\int d^6z~ j {\cal O}+\int d^6\bar{z}~ \bar{j}\bar{{\cal O}} \label{actionYM} \eea where we have 
set $q\equiv e^{i\pi\b}$ and in the following we choose $\b$ real so that $q\qb=1$. We have added 
to the classical action source terms for composite chiral operators generically denoted by ${\cal 
O}$ with $j$ ($\bar{j}$) (anti)chiral sources. The superfield strength $W_\a= i\Db^2(e^{-gV}D_\a 
e^{gV})$ is given in terms of a real prepotential $V$ and $\Phi_{1,2,3}$ contain the six scalars of 
the original ${\cal N}=4$ SYM theory organized into the ${\bf 3}\times \bf{ \bar 3}$ of $SU(3) 
\subset SU(4)$. We write $V=V^aT_a$, $\Phi_i=\Phi_i^a T_a$ where $T_a$ are $SU(N)$ matrices in the 
fundamental representation. In the following we use the notation ${\rm Tr} (T^aT^bT^c\dots)\equiv 
(abc\dots)$.

The $\beta$--deformation breaks the original $SU(4)$ R--symmetry to $U(1)_R$. However, 
besides the $Z_3$ symmetry associated to cyclic permutations of $(\Phi_1,\Phi_2,\Phi_3)$,
two extra
non--R--symmetry $U(1)$'s survive. Applying the $a$--maximization procedure \cite{IW} 
and the conditions of vanishing ABJ anomalies it turns out that $U(1)_R$ is the one which assigns 
the same R--charge $\o$ to the three elementary superfields, whereas the charges respect 
to the two non--R--symmetries $U(1)_1 \times U(1)_2$ can be chosen to be
$(\Phi_1,\Phi_2, \Phi_3) \rightarrow (0,1,-1)$ and $(-1,1,0)$, respectively. 

As discussed in \cite{FG,PSZ4}, at the quantum level the theory is superconformal invariant 
(and then finite) up to two loops if the coupling constants satisfy the following condition 
(vanishing of the beta functions)
\beq
 h \hb \left[ 1 - \frac{1}{N^2} \Big|q - \frac{1}{q} \Big|^2 \right]- g^2  = 0
\label{cond}
\eeq
In the large $N$ limit this condition reduces simply to $g^2 = h \hb$, independently of
the value of $q$. In \cite{RSS} the three loop correction to (\ref{cond}) has been also evaluated. 
Since it turns out to be suppressed for $N \to \infty$, the condition $g^2 = h \hb$ is the correct
condition for superconformal invariance in the planar limit up to three loops. 

At the superconformal fixed point of the theory we compute the anomalous dimensions for the 
class of non--protected operators 
\footnote{The choice of $\Phi_1$ and $\Phi_2$ superfields is totally arbitrary and we expect 
the operators ${\rm Tr }(\Phi_i^J \Phi_k)$, for any $i,k$ with $i \neq k$ to have similar quantum 
properties. We will comment on this point later on.}
\beq
{\cal O}_J = {\rm Tr }(\Phi_1^J \Phi_2) 
\label{OJ} 
\eeq
They are charged under $U(1)_1 \times U(1)_2$ with charges $(1,1-J)$. 

Using the equations of motion from the action (\ref{actionYM}) with $j=\bar{j}=0$ (from now on
we neglect factors of $e^{\pm gV}$ since they are not relevant to our purposes)
\beq 
\Db^2\Phib^a_3= -ih\Phi_1^b\Phi_2^c~[q(abc)-\frac{1}{q}(acb)] \label{eom} \eeq it is easy to see 
that \beq {\cal O}_J = \frac{i}{h[q - \frac{1}{q}]} \bar{D}^2 {\rm Tr}(\Phi_1^{J-1} \Phib_3) + 
\frac{1}{N} {\rm Tr}(\Phi_1^{J-1}) {\rm Tr}(\Phi_1 \Phi_2) \label{relation} 
\eeq 
As long as $J > 1$, in the large $N$ limit the operator ${\cal O}_J$ becomes descendent of the primary ${\rm 
Tr}(\Phi_1^{J-1} \Phib_3)$, whereas for finite $N$ the combination ${\cal O}_J - \frac{1}{N} {\rm 
Tr}(\Phi_1^{J-1}) {\rm Tr}(\Phi_1 \Phi_2)$ is descendent. The exceptional case $J=1$ corresponds to 
the chiral primary operator whose protection has been proven perturbatively in \cite{FG,PSZ4}. 

In the next Sections we will concentrate on the evaluation of the anomalous dimensions for
the ${\cal O}_J$ operators.

\section{The perburbative calculation}

In this Section we compute the anomalous dimension of ${\cal O}_J$ in (\ref{OJ}) perturbatively, up to two 
loops. For generic values of $J$ we perform the calculation in the large $N$ limit in order to
avoid dealing with mixing with multitrace operators. 
   
In order to compute anomalous dimensions we evaluate one--point correlators 
$\langle {\cal O}_J e^{S_{int}} \rangle$ where $S_{int}$ is the sum of the interaction terms in 
(\ref{actionYM}).
Divergent contributions proportional to the operator itself 
are removed by a multiplicative renormalization which in dimensional regularization reads
\beq
{\cal O}_J^{(bare)} \equiv  {\cal O}_J \left( 1 + \sum_{k=0}^{\infty} \frac{a_k(\l, q, N )}{\e^k} \right) 
\equiv Z {\cal O}_J
\eeq
where we have introduced the 't Hooft coupling $\l = \frac{g^2N}{4\pi^2}$. 
There is no dependence on the $h$ coupling since we are at the superconformal point where $h$ can be 
expressed in terms of the other couplings through the condition of vanishing beta functions. 

The anomalous dimension is then given by
\beq
\g \equiv 2\l \frac{da_1(\l, q, N)}{d\l}  
\label{andim}
\eeq
Therefore, at any order it is easily read from the simple pole divergence. 

We perform perturbative calculations in a superspace setup by following closely the procedure used 
in \cite{PSZ1, PSZ2, PSZ3, PS, PSZ4} (we refer to those papers for conventions and technical details). 
After D--algebra supergraphs are reduced to ordinary Feynman diagrams which we evaluate in momentum 
space. We work in dimensional regularization, $d = 4-2\e$, and minimal subtraction scheme.  

From the action (\ref{actionYM}) quantized in the Feynman gauge we read the superfield propagators
($z \equiv (x, \th,\thb)$)
\bea
&& \langle V^a(z_1) V^b(z_2) \rangle = ~~\d^{ab} \frac{1}{(x_1 - x_2)^2} \d^{(4)}(\th_1 - \th_2)
\nonumber \\
&&  \langle \Phi_i^a(z_1) \Phib_j^b(z_2) \rangle  = ~- \d_{ij}\d^{ab} 
\frac{1}{(x_1 - x_2)^2} \d^{(4)}(\th_1 - \th_2)
\label{prop}
\eea
and the three--point vertices
\bea
&& (\Phi \Phi \Phi)_{{\rm vertex}}  \rightarrow ~~~~~~
ih \Phi_1^a \Phi_2^b \Phi_3^c \left[ q (abc) - \frac{1}{q} (acb) \right]
\nonumber \\
&& (\Phib \Phib \Phib)_{{\rm vertex}}  \rightarrow ~~~
-i\hb \Phib_1^a \Phib_2^b \Phib_3^c \left[ \qb (acb) - \frac{1}{\qb} (abc) \right]
\nonumber \\
&& (\Phib V \Phi)_{{\rm vertex}} \rightarrow ~~~~~~ g \Phib_i^a V^b \Phi_i^c ~\left[ (abc) -(acb) \right]
\label{vertices}
\eea

At the lowest order the only contribution to the one--point function for the operator ${\cal O}_J$
is the one given in Fig. 1 where, using the notation introduced in \cite{GMR},  
the horizontal bold line indicates the spacetime point where the operator is inserted. 
\vskip 18pt
\noindent
\begin{minipage}{\textwidth}
\begin{center}
\includegraphics[width=0.40\textwidth]{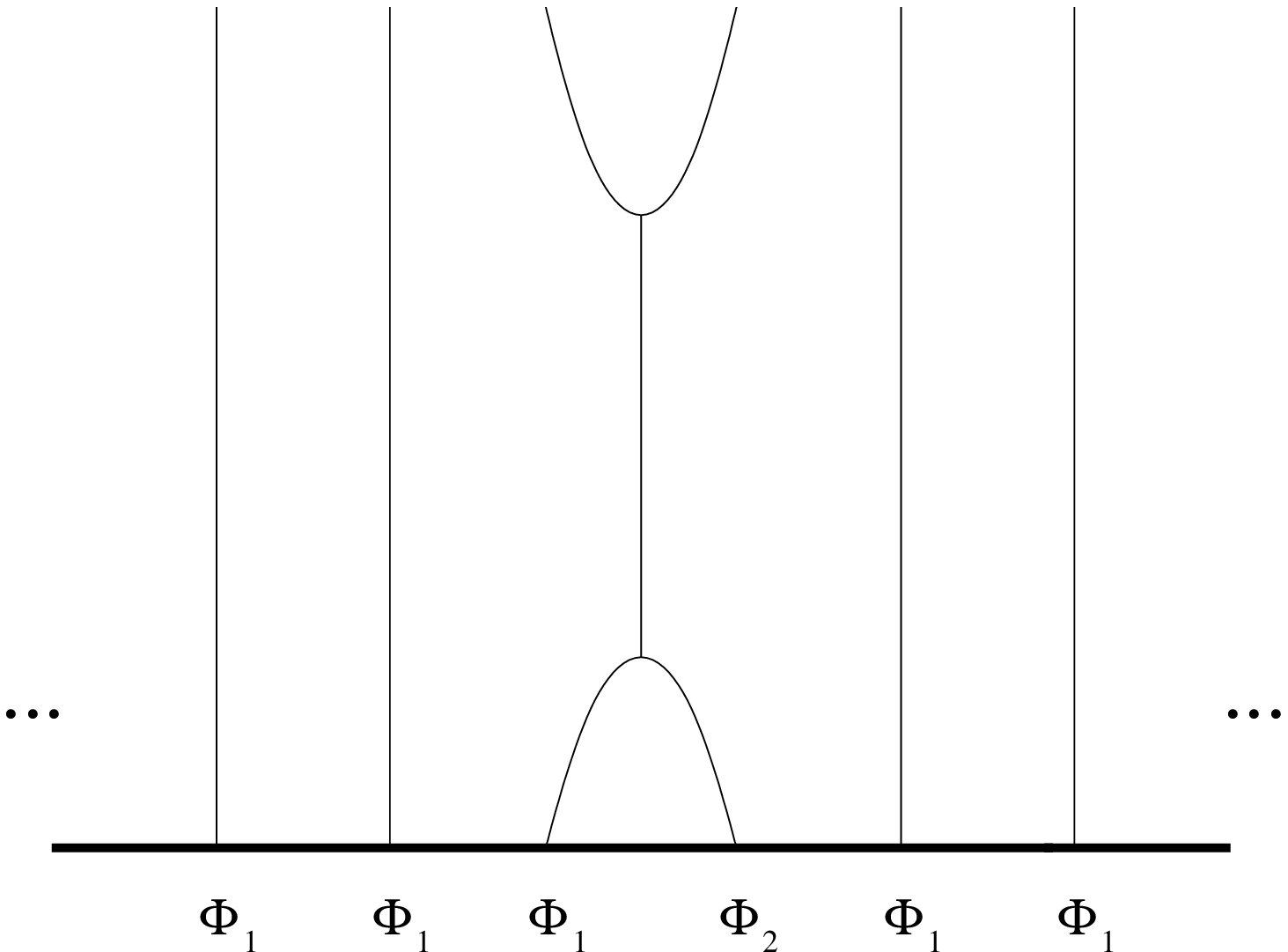}
\end{center}
\begin{center}
{\small{Figure 1: One--loop contribution to the ${\cal O}_J$ anomalous dimension}}
\end{center}
\end{minipage}

\vskip 20pt The corresponding contribution is proportional to the self--energy integral \beq I_1 
\equiv \int d^n k \frac{1}{k^2 (p-k)^2} \sim \frac{1}{(4\pi)^2} \frac{1}{\e} \label{selfenergy} 
\eeq Evaluating the color factor, the combinatorics and taking into account a minus sign from 
D--algebra we obtain \beq {\rm Diagram ~1} ~~~\rightarrow ~~~- \frac{1}{\e} ~ | q - \frac{1}{q} |^2 
\frac{|h|^2 N}{(4\pi)^2} \eeq Using the one--loop superconformal condition in the planar limit 
($g^2 = |h|^2$) and the definition (\ref{andim}) we immediately find the one--loop anomalous 
dimension \beq \g^{(1)} = \frac12 \Big| q - \frac{1}{q} \Big|^2 \l \label{gamma1} \eeq

At two loops (order $\l^2$) the diagrammatic contributions are drawn in Fig. 2. 
\vskip 18pt
\noindent
\begin{minipage}{\textwidth}
\begin{center}
\includegraphics[width=0.80\textwidth]{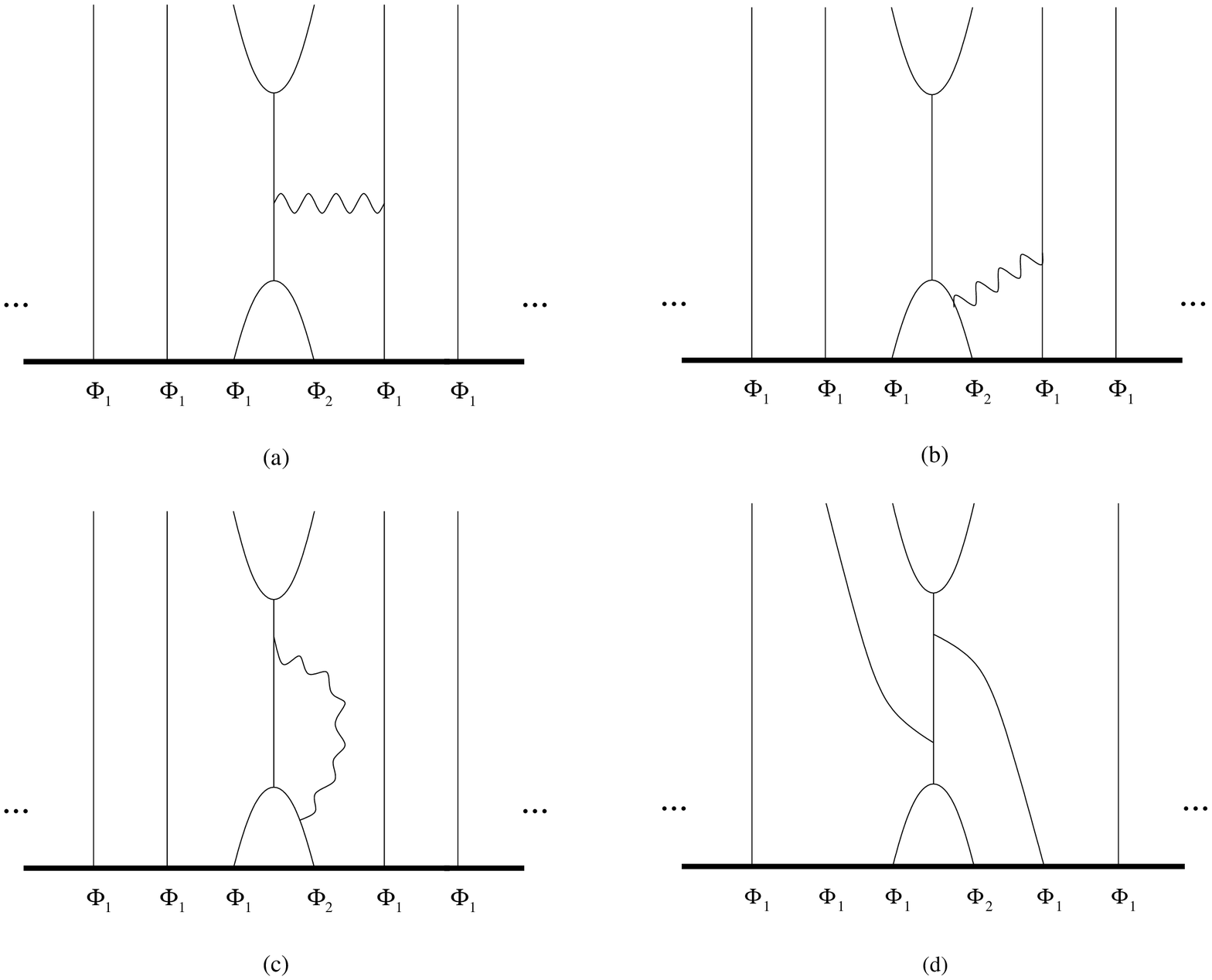}
\end{center}
\begin{center}
{\small{Figure 2: Two--loop contributions to the ${\cal O}_J$ anomalous dimension}}
\end{center}
\end{minipage}

\vskip 20pt
Performing the D--algebra we reduce all the diagrams to ordinary Feynman diagrams 
containing the loop structure as in Fig. 3. 

\vskip 18pt
\noindent
\begin{minipage}{\textwidth}
\begin{center}
\includegraphics[width=0.50\textwidth]{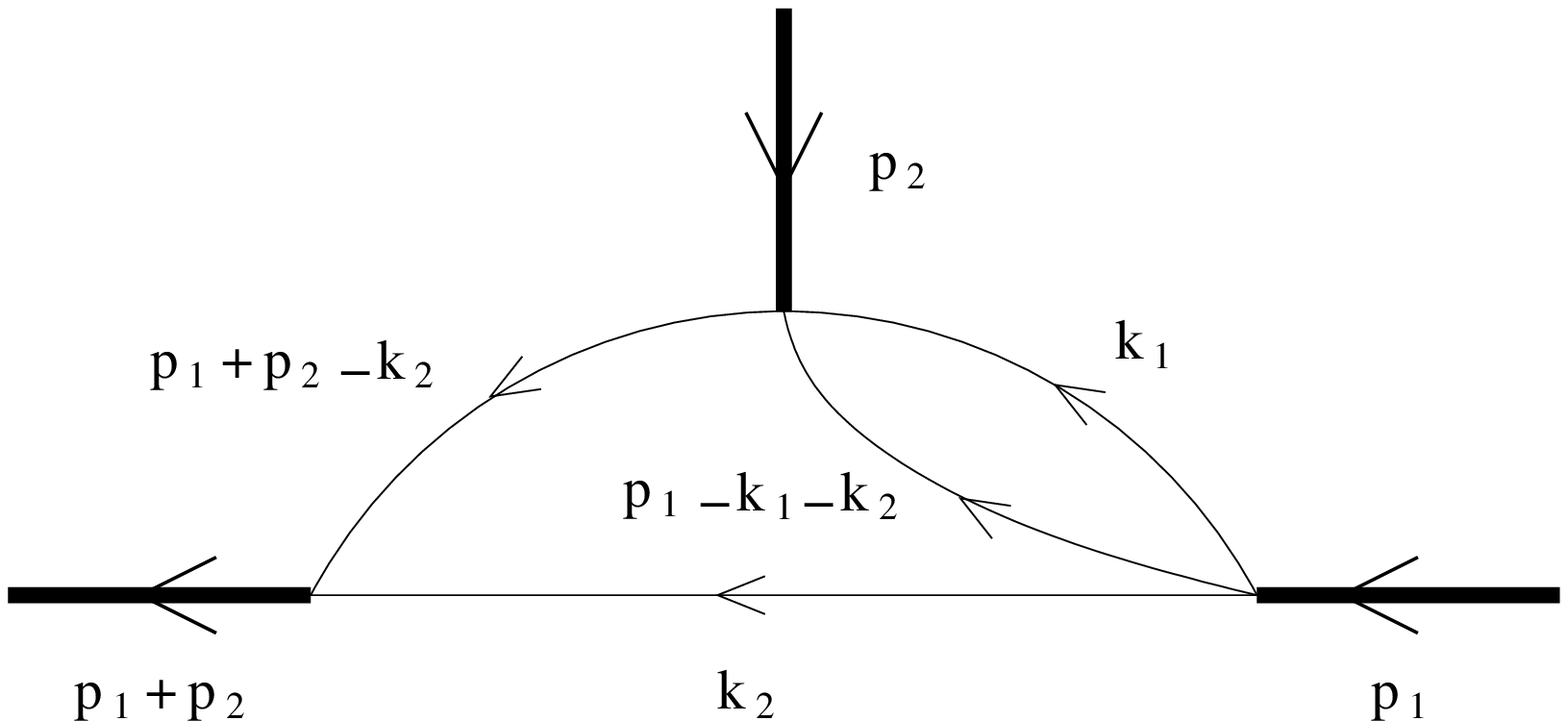}
\end{center}
\begin{center}
{\small{Figure 3: The two--loop Feynman diagram}}
\end{center}
\end{minipage}

\vskip 20pt

The associated momentum integral is
\beq
I_2 \equiv \int d^n k_1 d^n k_2 \frac{1}{k_1^2 (p_1 - k_1 - k_2)^2 k_2^2 (p_1+p_2-k_2)^2}  
\eeq
As long as we are only concerned with UV divergences we can safely set one of the
external momenta to zero. Thus the graph is easily evaluated being proportional
to two nested self--energies. We obtain (in the G-scheme) 
\beq
I_2 \sim \frac{1}{(4\pi)^4}~ \frac{1}{2\e^2} (1 + 5 \e) \frac{1}{(p^2)^{2\e}}
\eeq
where we have kept only divergent terms. Performing the subtraction of the subdivergence 
we finally have
\beq
\left[ I_2 \right]_{sub} \sim \frac{1}{(4\pi)^4} ~\left[ - \frac{1}{2\e^2} + \frac{1}{2\e} \right]
\label{I2}
\eeq
Computing the combinatorics, the color factors and taking into account minus signs from the vector
propagator we find that the factors in front of (\ref{I2}) for the various diagrams are
\bea
&& (2a) ~~~ \rightarrow ~~~ - 2 (q - \frac{1}{q}) (\qb - \frac{1}{\qb}) g^2 |h|^2 N^2 
\nonumber \\
&& (2b) ~~~ \rightarrow ~~~ ~~ 2 (q - \frac{1}{q}) (\qb - \frac{1}{\qb}) g^2 |h|^2 N^2
\nonumber \\
&& (2c) ~~~ \rightarrow ~~~  ~~ 2 (q - \frac{1}{q}) (\qb - \frac{1}{\qb}) g^2 |h|^2 N^2
\nonumber \\
&& (2d) ~~~ \rightarrow ~~~ - (q - \frac{1}{q}) (\qb - \frac{1}{\qb}) \left( \frac{q}{\qb} + 
\frac{\qb}{q} \right)|h|^4 N^2
\eea
Summing all the contributions, using the planar superconformal condition $ |h|^2 = g^2$ and 
the definition (\ref{andim}), we find
\beq
\g^{(2)} = -\frac18 \Big| q  - \frac{1}{q} \Big|^4 \l^2
\label{gamma2}
\eeq
We observe that the diagrams contributing to the anomalous dimensions for our operators are
exactly the same as the ones for BMN operators in ${\cal N}=4$ SYM in the planar limit 
\cite{BMN,GMR}. In fact, up to this order the calculation is exactly the same under the
formal identification $|q - \frac{1}{q}|^2 \leftrightarrow -(e^{i\phi} +e^{-i\phi}-2)$,
where $\phi$ is the phase of BMN operators \cite{BMN,GMR,SZ}. We expect that the same
pattern will persist at any order in perturbation theory. In particular, as in 
the BMN case, the graphs relevant for the calculation are only the ones where the
interactions are close to the ``impurity'' $\Phi_2$: at $L$--loop order the 
interactions may involve at most the $\Phi_1$ lines which are $L$--steps far away from
the impurity. As an important consequence, in the large $J$ limit the anomalous 
dimensions do not grow with $J$.

To close this Section we note that the result we have found for the anomalous dimensions of
the operators ${\rm Tr}(\Phi_1^J \Phi_2)$ at large $N$ are actually valid for any operator of the
form ${\rm Tr}(\Phi_i^J \Phi_k)$ with $i \neq k$. In fact the superpotential is invariant 
under cyclic permutation of
$(\Phi_1, \Phi_2,\Phi_3)$, and in addition it  becomes invariant if non--cyclic exchanges 
of fields are accompanied by 
\beq
q \rightarrow -\frac{1}{q}
\label{q}
\eeq
Since the anomalous dimensions are proportional to 
powers of the effective coupling $\a \equiv \l \Big| q  - \frac{1}{q} \Big|^2$ which is invariant 
under (\ref{q}) we conclude that the result is valid for any operator of the form
${\rm Tr}(\Phi_i^J \Phi_k)$, $i \neq k$.

\section{The exact anomalous dimensions} 

Motivated by the formal correspondence of the previous calculation with the BMN case, 
in this Section we are going to compute the {\em exact} anomalous
dimensions in the large $N$, large $J$ limit by using the procedure introduced in 
\cite{SZ} for BMN operators. In the context of $\beta$--deformed theories the same procedure 
has been applied to the class of BMN operators \cite{NP}.

We concentrate on the operator ${\cal O}_{J+1}$ which, as follows from eq. (\ref{relation}), 
in the planar limit satisfies 
\beq
\bar{D}^2 {\cal U}_{J} = - ih [q - \frac{1}{q} ] {\cal O}_{J+1}
\label{Oop}
\eeq
where we have defined
\beq
{\cal U}_{J} \equiv {\rm Tr} (\Phi^J_1\Phib_3) 
\label{Uop}
\eeq
As already noticed, this shows that the ${\cal O}_{J+1}$ operators are 
descendants of the ${\cal U}_{J}$ ones.
Being part of the same superconformal multiplet they share the renormalization properties, i.e. they will 
have the same scaling dimension and the same perturbative corrections to their overall
normalization. Moreover since ${\cal U}_{J}$ is not a Konishi-like operator 
it is not affected by the Konishi anomaly.

As discussed in details in \cite{SZ}, in any $N=1$ superconformal field theory the two--point function 
for a {\em primary} operator ${\cal A}_{s,\bar{s}}$ is fixed \cite{O} and given by ($z \equiv (x, \th, \bar{\theta})$)
\bea
&&<{\cal{A}}_{(s,\bar{s})}(z)\bar{{\cal{A}}}_{(s,\bar{s})}(z')>
=f_{{\cal{A}}}(g^2,N,h,\bar{h})
\left\{\frac{1}{2}D^\a \bar D^2 D_\a +  \frac{w}{4(\Delta_0+\g)}
[ D^\a,\bar D^{\ad} ] i\pa_{\a\ad} \right.~~~~~~\nonumber\\
&&~~~~~\nonumber\\
 &&~~~~~~~~~~~~~~~~~~~~~~~~~~~~~~~\left.+ 
\frac{(\Delta_0+\g)^2 + w^2 -2(\Delta_0+\g)}{4(\Delta_0+\g) (\Delta_0 +\g- 1)}\square\right\}
\frac{\delta^4(\theta-\theta')}{|x-x'|^{2(\Delta_0+\g)}}
\label{niceformula}
\eea
where $\D_0 = s + \bar{s}$ is the tree--level dimension of the operator, $\o = s - \bar{s}$ is
its R--symmetry charge \footnote{We assume $\o$ not to renormalize. In fact, once the R--symmetry
of the elementary fields is fixed by requiring the exact R--symmetry of the superpotential,
any composite operator has a fixed charge given by the sum of the charges of its elementary 
constituents.} and $\g$ is the exact anomalous dimension. 

The relation (\ref{niceformula}) can be straightforwardly applied to our primary operators
${\cal U}_J$. 
The analysis of the two-point correlator for the ${\cal O}_{J}$'s 
is somewhat subtler since, as we see from eq. (\ref{Oop}), these chiral operators are  
not primaries and in principle the relation (\ref{niceformula}) cannot be applied to
their correlators.
However, as we are going to show, in the large $J$ limit these operators turn out to 
behave as CPO's and (\ref{niceformula}) can be safely used.\\ 

To this end we remind that in general given a {\em chiral} operator ${\cal A}$, 
the condition for the operator to be non--protected 
(anomalous dimension acquired) is equivalent to the condition that its chiral
nature is not maintained under superconformal transformations, i.e. 
$\bar{D}(\d_{\bar{S}} {\cal A}) \sim \{\bar{S},\bar{D}\}{\cal A} \neq 0$ (see for instance
\cite{CW}).
In fact, writing schematically the superconformal algebra relation for a scalar 
operator as $\{\bar{S},\bar{D}\} = \Delta - \omega$, we have
\beq
\{\bar{S},\bar{D}\} {\cal A} = (\Delta  - \omega){\cal A} 
= \left[ (\D_{0} + \g) - \o \right] {\cal A} = \g {\cal A}
\eeq 
where we have used $\D = \D_{0} +\g$ and for a chiral operator $\o = \D_{0}$. Therefore if 
$\g \neq 0$, $ \bar{S} {\cal A}$ is not chiral anymore. Viceversa, if $\{\bar{S},\bar{D}\} {\cal A}
=0$, then $\bar{S}{\cal A}$ is still chiral and the dimension is protected by the 
well--known condition $\Delta = \omega$.

An alternative proof goes through the simple observation that the conditions
\beq
s+\bar{s}=\D_{0}+\g \qquad\qquad s-\bar{s}=\D_{0}
\eeq
imply
\beq
s=\D_{0}+\frac{\g}{2}\qquad\qquad \bar{s}=\frac{\g}{2} 
\eeq
The appearance of $\bar{s} \neq 0$ signals the lack of chirality of the quantum
operator. 

We now apply the previous argument to our operators $O_J$ to prove that in the 
large $N$, large $J$ limit the violation of chirality is suppressed and 
they behave as CPO's. In the limit of large R--symmetry $\o = J$ it is more natural to
consider
\beq
\frac{1}{J} \{\bar{S},\bar{D}\} {\cal O}_J = \left( \frac{\D_{0}+\g}{J} -1 \right){\cal O}_J
= \frac{\g}{J} {\cal O}_J
\label{Jlarge}
\eeq
As discussed in the previous Section, at any fixed order in
perturbation theory the anomalous dimension $\gamma$ does not grow with $J$. 
It follows that in the large $J$ limit 
the r.h.s. of eq. (\ref{Jlarge}) is suppressed and the operator behaves as a chiral primary. 
In particular, in this limit it is 
consistent to apply eq. (\ref{niceformula}) for the evaluation of its two--point function. 

Supported by these considerations we can now proceed exactly as in \cite{SZ} and 
find
\bea
&&<\bar{D}^2{\cal{U}}_J(z)D^2\bar{{\cal{U}}}_J(z')>=\nonumber\\
&&~~~~~~\nonumber\\
&&~~~=\frac{N^{J+1}}{(4\p^2)^{J+1}}
f ~\bar{D}^2\left\{\frac{1}{2}D^\a \bar D^2 D_\a + \frac{J-1}{4(J+1+\g)}
[ D^\a,\bar D^{\ad} ] i\pa_{\a\ad} \right.~~~~~~\nonumber\\
&&~~~~~\nonumber\\
 &&~~~~~\left.+ 
\frac{(J+1+\g)^2 + (J-1)^2 -2(J+1+\g)}{4(J+1+\g) (J +\g)}\square\right\}
D^2\frac{\delta^4(\theta-\theta')}{|x-x'|^{2(J+1+\g)}}\nonumber\\
&&~~~= \frac{N^{J+1}}{(4\p^2)^{J+1}}
f~ (\g^2+2\g)\bar{D}^2 D^2\frac{\delta^4(\theta-\theta')}
{|x-x'|^{2(J+2+\g)}}
\label{corrUU}
\eea
and
\beq
<{\cal{O}}_{J+1}(z)
\bar{{\cal{O}}}_{J+1}(z')>=\frac{N^{J+2}}{(4\p^2)^{J+2}}
f \bar{D}^2D^2 \frac{\delta^4(\theta-\theta')}{|x-x'|^{2(J+2+\g)}}
\label{corrOO}
\eeq
where $f$ is the common normalization function not fixed by superconformal
invariance. 

From the relation (\ref{Oop}) the two correlators are related by
\beq
\langle\bar{D}^2  {\cal U}_J (z)  D^2 \bar{\cal U}_J (z')\rangle  ~=~ 
|h|^2 \Big| q - \frac{1}{q} \Big|^2 \langle {\cal O}_{J+1} (z) 
\bar{\cal O}_{J+1} (z')\rangle
\label{OU}
\eeq
Therefore, inserting in (\ref{OU}) the explicit expressions (\ref{corrUU}) and
(\ref{corrOO}) we end up with an algebraic equation 
\beq
\g^2 + 2\g = |h|^2 \Big| q - \frac{1}{q} \Big|^2 \frac{N}{4\pi^2}
\eeq
 which allows to find the exact expression for the anomalous dimensions
\bea
\g &=& -1 + \sqrt{1+ |h|^2 \Big| q - \frac{1}{q} \Big|^2 \frac{N}{4\pi^2}}
\nonumber \\
&= & \frac12 |h|^2 \Big| q - \frac{1}{q} \Big|^2 \frac{N}{4\pi^2} ~-~
\frac18 |h|^4 \Big| q - \frac{1}{q} \Big|^4 \frac{N^2}{(4\pi^2)^2} ~+ ~\cdots
\label{andimexact}
\eea
Up to the second order this expression coincides with the perturbative
results obtained in the previous Section.

We note that our operators ${\cal O}_J$ can be thought as dual to the
0--modes of the BMN sector considered in \cite{LM,NP,FRT1,FRT2}. Formula 
(\ref{andimexact}) is in agreement with the results presented in 
those papers for the spectrum of the 0--modes.

\section{The superconformal condition at large $N$}

In the previous Section, exploiting the superconformal invariance of the theory and its
equations of motion we have shown that the exact anomalous dimension for the ${\cal O}_J$
operator can be written as
\beq
\g = -1 + \sqrt{1+ 2 \g^{(1)}}
\label{andimexact2}
\eeq
where $\g^{(1)}$ is the one--loop anomalous dimension. A direct calculation 
provides an expression for $\g^{(1)}$ proportional to $|h|^2$. As discussed in Section 3,
at this order we are allowed to use the planar superconformal condition $|h|^2 =g^2$
to re-express $\g^{(1)}$ in terms of $g^2$ only (see eq. (\ref{gamma1})). Now if in (\ref{andimexact2}) 
we use $\g^{(1)}$ given
in terms of $g^2$ and expand the square root, we obtain a perturbative formula for $\g$
which agrees with the actual perturbative calculation only if the condition  $|h|^2 =g^2$
is valid at any order.

Motivated by this observation we are led to conjecture that in the large $N$ limit 
the condition $|h|^2 =g^2$ is indeed the correct condition for superconformal 
invariance at any order in perturbation theory. Direct confirmations of this 
conjecture can be found in the literature up to order $g^6$ \cite{FG, PSZ4, RSS}. 
Now we give an argument to prove that this is true to all orders.

We remind that in ${\cal N}=1$ supersymmetric theories the superconformal invariance
condition (i.e. vanishing of beta functions) can be expressed as the vanishing of 
the anomalous dimensions of the elementary superfields \cite{GRS,S,LS,JJ}. Therefore, in order to study
superconformal invariance, it is sufficient to focus on the  
divergent corrections to the propagators of the elementary fields. 

In the $\b$--deformed theory we consider a generic $L$--loop diagram contributing to 
the propagator of the $\Phi_i$ superfield. The crucial observation is
the following: If we prove that at the planar level, as long as $q\qb=1$, this diagram 
does not depend on $q$, then we are sure that $|h|^2 =g^2$ is the exact solution 
of the superconformal invariance equations. In fact, if it is independent of $q$,   
the corresponding perturbative contribution is the same for any deformed theory,
independently of the choice of the $q$--deformation. In particular, it is the same 
for any deformed theory ($q \neq 1$) and for the underformed one ($q=1$). 
Focusing on the undeformed case we can conclude that $|h|^2 =g^2$ is 
the exact condition for the planar superconformal invariance, since 
$q=1$ and $|h|^2 =g^2$ bring us back to the ${\cal N}=4$ case which is known to
be exactly superconformal. The independence of the perturbative corrections on $q$
allows to extend this statement to any deformed theory.

To conclude the proof we need to show that the contribution from a generic self--energy
planar diagram never depends on $q$. We can focus on diagrams
containing only matter vertices because adding vector propagators
cannot introduce any $q$-dependence. We exploit the formal analogy between the 
deformed theory and noncommutative (nc) field theory. As observed in \cite{LM} the
deformed potential can be written as 
\beq
ih \int d^6z ~{\rm Tr}(\Phi_1 \ast [\Phi_2 , \Phi_3]_{\ast}) ~+~ {\rm h.c.}
\eeq
where 
\beq
f \ast g = e^{i \pi \b Q^{(i)}_f M_{ij} Q^{(j)}_g} f \cdot g ~,
\label{star}
\eeq
$Q^{(i)}$, $i=1,2$ being the non--R--symmetry $U(1)_1 \times U(1)_2$ charges and
$M$ the antisymmetric matrix $ \left( \begin{matrix} {0 & 1 \cr
                                 -1 & 0}
                  \end{matrix} \right)$.
When drawing a Feynman diagram we can consider the flow of the charges inside 
the diagram. Observing that the charges are conserved at any vertex and propagate
through the straight lines we can formally identify them with the ordinary momenta
in noncommutative diagrams. A known property of planar diagrams in nc field theory
is that the star product phase factors dependent on the loop momenta cancel out (for a proof
see \cite{F,IIKK,MSV}) and only an overall phase depending on the external 
momenta survives. In our case, exploiting the formal identification of charges
with momenta, we can use the same arguments to conclude that any planar diagram
will have a phase factor from (\ref{star}) depending only on the configuration of the external 
charges. In the particular case of self--energy diagrams the overall phase is zero 
since $\Phi_i$ and $\Phib_i$ have equal (but opposite) charges. 
In other words, any self--energy planar diagram always contains
an equal number of $q= \frac{1}{\qb}$ and $\qb= \frac{1}{q}$ vertices.
This concludes the proof of the $q$ independence of perturbative self--energy corrections.

We state that in the $N\rightarrow\infty$ limit the exact condition for
superconformal invariance is simply $|h|^2 =g^2$. Therefore, the theory described by the
action
\bea
&& S =\int d^8z~ {\rm Tr}\left(e^{-gV} \Phib_i e^{gV} \Phi^i\right)+
\frac{1}{2g^2}
\int d^6z~ {\rm Tr} W^\a W_\a
\nonumber\\
&&+ig \int d^6z~ {\rm Tr}( e^{i\pi\b} \Phi_1 \Phi_2 \Phi_3 - e^{-i\pi\b} \Phi_1 \Phi_3 \Phi_2 )
+ i g \int d^6\bar{z}~ {\rm Tr} ( e^{i\pi\b} \Phib_1 \Phib_2 \Phib_3 - e^{-i\pi\b} 
\Phib_1 \Phib_3 \Phib_2 )
\nonumber  \\
&&~~~
\label{confaction}
\eea 
represents a ${\cal N}=1$ superconformal invariant theory for any value of $\b$ real.

\section{Conclusions}

For the $SU(N)$, $\b$--deformed ${\cal N}=4$ SYM theory we have considered the particular class of 
operators ${\cal O}_{J} = {\rm Tr}(\Phi_1^J \Phi_2)$. We have computed perturbatively their
anomalous dimensions up to two loops. The calculation has been performed in the large $N$ 
limit in order to avoid mixing with multi-trace operators. Exploiting the techniques introduced
in \cite{SZ} for the BMN operators we have evaluated their {\em exact} anomalous dimensions
in the large $N$, large $J$ limit. In this limit the exact expression for the anomalous dimension 
depends on the deformation parameter $q$ only through the combination $|q - \frac{1}{q}|^2$ and it is
then invariant under $q \rightarrow -\frac{1}{q}$. Observing that in $\b$--deformed theory exchanging 
$q \rightarrow -\frac{1}{q}$  amounts to exchange $\Phi_i \leftrightarrow \Phi_j$, for any $i=1,2,3$, 
$i \neq j$, we may conclude that in the ${\cal O}_J$ sector, in the large $N$, large $J$ 
limit there is an enhancement of the $SU(3)$ symmetry and all the operators of the form 
 ${\rm Tr}(\Phi_i^J \Phi_k)$, $i\neq k$ renormalize in the same way.

A comparison between the exact result and the perturbative calculation suggests that the
condition $|h|^2 = g^2$, which up to three--loops guarantees the superconformal invariance of
the theory in the planar limit, is actually sufficient for the exact invariance  
for $N \to \infty$. Indeed, we have given a direct proof at any order in
perturbation theory. The main result of our paper is that the action in 
(\ref{confaction}) is superconformal invariant at the quantum level without additional conditions 
on the couplings. 
In the context of the AdS/CFT correspondence \cite{M,GKP,W} this is the theory whose strong coupling
phase is described by the supergravity dual found in \cite{LM}.  
The ${\cal O}_{J}$ sector of this theory for large $J$ 
shares many similarities with the BMN sector of the ${\cal N}=4$ theory in the pp--wave limit.
This opens the possibility for these operators to be dual to superstring states in some 
particular sector of the theory. 
   
It is interesting to consider the extension of our calculations 
to the case of $\b$ complex ($q \qb \neq 1$). In this case the condition 
for superconformal invariance up to three loops,
in the planar limit becomes 
\beq
\frac12 |h|^2 \left( q\qb + \frac{1}{q\qb} \right) = g^2
\label{condcomplex}
\eeq
When (\ref{condcomplex}) holds it is easy to see that the perturbative anomalous dimension 
still coincides with the expansion of the exact result (\ref{andimexact}). 
As for the case of 
$\b$ real consistency of the perturbative calculation with the exact result would
suggest that the condition (\ref{condcomplex}) should be valid at any order. However, 
the proof we have presented in Section 5 makes repeated use of the requirement
$q\qb=1$ and cannot be immediately extended to the more general case. A different 
procedure should be found to prove or disprove that the one--loop condition 
(\ref{condcomplex}) is sufficient to insure the exact conformal invariance even in 
the case of $\b$ complex.   
 
Generalizations of the present results to other deformed theories are presently under 
investigation and will be reported in \cite{MPPSZ}.

\vspace{1.5cm}

\section*{Acknowledgements}

\noindent This work has been supported in
part by INFN, PRIN prot. 2003023852\_008 and the European Commission RTN 
program MRTN--CT--2004--005104.

\end{document}